\documentclass[prl,twocolumn]{revtex4}

\usepackage{graphicx}

\newcommand{\ra}{\rangle}
\newcommand{\la}{\langle}
\newcommand{\ua}{\uparrow}
\newcommand{\da}{\downarrow}
\begin{document}
\title{Entanglement of $\eta $-pairing state with
off-diagonal long-range order
}

\author{Heng Fan$^1$, Seth Lloyd$^2$}
\affiliation{
$^1$Quantum computation and information project,
ERATO, Japan Science and Technology Agency,\\
Daini Hongo White Building 201, Hongo 5-28-3, Bunkyo-ku, 
Tokyo 113-0033, Japan.\\
Electrical Engineering Department, UCLA, Los Angeles, CA 90095, USA\\
$^2$Department of Mechanical Engineering, MIT, Cambridge, MA 02139, USA 
}
%\maketitle
                                                        
\begin{abstract}
Off-diagonal long-range order (ODLRO) is 
a quantum phenomenon not describable in 
classical mechanical terms.
It is believed to be one characteristic
of superconductivity.
The quantum state constructed by $\eta $-pairing 
which demonstrates ODLRO is
an eigenstate of the three-dimensional Hubbard model.
Entanglement is a key concept of the quantum information
processing and has no classical counterpart. 
We study the entanglement property of $\eta $-pairing quantum
state. The concurrence is a well-known
measure of quantum entanglement. We show that
the concurrence of entanglement between one-site
and the rest sites is exactly the correlation function
of the ODLRO for the $\eta $-pairing state in the
thermodynamic limit.   
So, when the $\eta $-pairing state is entangled, 
it demonstrates ODLRO and is thus in superconducting phase, 
if it is a separable state,
there is no ODLRO. 
In the thermodynamical limit, the entanglement between
$M$-site and other sites of the $\eta $-pairing
state does not vanish.
Other types of ODLRO of $\eta$-pairing state are presented. We show that
the behavior of the ODLRO correlation functions
is equivalent to that of the entanglement of the $\eta$-pairing state.
The scaling of the entropy of the entanglement for the
$\eta $-pairing state is studied.
\end{abstract}
\maketitle

Quantum entanglement plays a central role in quantum
information and quantum computation. 
It is regarded as a resource for processing quantum
information\cite{BD,L}.
With quantum entanglement,
we can perform quantum teleportation\cite{BBCJPW}, 
super-dense coding\cite{BW},
quantum cryptography\cite{E}, universal quantum computation by
only measurements\cite{GC}.
It can enhance the classical capacity
and quantum communication\cite{BSST,L1}, enhance the
accuracy of clock synchronization\cite{GLM}, etc.

On the other hand, quantum entanglement may also be regarded as
an important order parameter in quantum phase transitions\cite{S}.
Recent results showed that the pair-wise quantum entanglement
between nearest neighbor sites and next nearest neighbor
sites of the ground state of $XY$ spin chain
displays a peak either
near or at the critical point of quantum phase 
transition\cite{OAFF}. Similar phenomenon also appears when
the entanglement between one site and the rest of other sites
is studied\cite{ON}. It is also pointed out that
entanglement of the ground state 
of $XXZ$ and $XY$ spin chains at critical
point is analogous to that of entropy in conformal 
field theories\cite{VLRK}. Some related works were
also done for various models\cite{K,ABV,VMC,PP,V}. 
Recent experiments showed that entanglement can contribute
significantly to the bulk susceptibility\cite{GRAC}.

ODLRO is an important concept in condensed matter physics\cite{Y,Y1}.  
It is proved that ODLRO implies both
Meissner effect and the quantization of 
magnetic flux, which are the basic
characteristic properties of superconducting states\cite{Y,NSZ}.
Yang argued that since off-diagonal elements have no classical
analog, the off-diagonal long-range order is quantum phenomenon
not describable in classical mechanical terms\cite{Y}.
From quantum information theory,
it is presently obvious that quantum entanglement
is quantum mechanical and has no classical counterpart.
And also entangled state may be shared by spatially
separated parties and thus can have long-range correlation.
Consequently, superconductors may also be characterized by
the existence of quantum entanglement. And 
the property of quantum entanglement could be the hidden reason
that ODLRO can characterize the superconductivity.
We may also argue that entanglement may be one basic quantity
to describe different systems. So, it is necessary
to explore the property of 
entanglement for various quantum systems.

Most of the previous results about entanglement in
quantum critical phenomena are for one-dimensional
systems. One reason is that higher-dimensional
system is difficult to solve exactly, for
example, Hubbard model and most of the spin models
unless some approximations are chosen.
However, in three dimensions (also in one and
two dimensions), it is well-known that Hubbard model has an eigenstate
with $\eta$-pairing\cite{Y1}. It is argued that this 
state is metastable.      
We know that this eigenstate with $\eta $-pairing
possesses ODLRO. 
And also this state is symmetric and relatively
simple, the entanglement
of this state can be analyzed explicitly. 
In this paper, we will study the entanglement 
of the $\eta $-pairing quantum state.  

The Hamiltonian of the Hubbard model is as follows:
\begin{eqnarray}
H&=&-\sum _{\sigma ,<j,k>}
(c^{\dagger }_{j,\sigma }c_{k,\sigma }
+c^{\dagger }_{k,\sigma }c_{j,\sigma })
\nonumber \\
&&+U\sum _{j=1}^L(n_{j\ua }-\frac 12)(n_{j,\da }-\frac 12),
\nonumber
\end{eqnarray}
where $\sigma =\ua ,\da $, and $j, k$ are
nearest neighboring sites, $n_{j,\sigma }=c^{\dagger }_{j,\sigma }
c_{j,\sigma }$ are number operators. 
$c^{\dagger }_{j,\sigma }$ are standard fermion operators with
anticommutation relations given by 
$\{c^{\dagger }_{j,\sigma }, c_{k,\sigma '}\}=
\delta _{j,k}\delta _{\sigma ,\sigma '}$.
We assume the lattice under consideration is three dimensions,
and the total number of lattice sites is $L$.
The $\eta $-pairing operators at lattice site 
$j$ are defined as
$\eta _j=c_{j,\ua}c_{j,\da}, 
\eta _j^{\dagger }=c_{j,\da}^{\dagger }c_{j,\ua}^{\dagger },
\eta _j^{z}=-\frac {1}{2}n_j+\frac 12$.
These operators  form a $SU(2)$
algebra as showed by the relations
$[\eta _j,\eta _j^{\dagger }]=2\eta _j^z,
[\eta _j^{\dagger },\eta _j^z]=\eta _j^{\dagger },
[\eta _j,\eta _j^z]=-\eta _j$.
The $\eta $ operators are defined as
$\eta =\sum _{j=1}^L\eta _j, \eta ^{\dagger }
=\sum _{j=1}^L\eta _j^{\dagger }$.
Yang pointed out that the following quantum state is an
eigenstate of the Hubbard model\cite{Y1} 
\begin{eqnarray}
|\Psi \ra =(\eta ^{\dagger })^N|vac\ra .
\label{state}
\end{eqnarray}
This quantum state is not only an eigenstate of the
Hubbard model, but also an eigenstate of other
models which are exactly solvable by Bethe ansatz method
in one dimension\cite{EKS,F}.
In particular, it is the ground state for the model
in \cite{EKS} for a special case. 
We should also point out that the quantum state
(\ref{state}) actually is not tied to any specific model.
And it can be in any dimensions and with any lattice 
configurations.
The ODLRO of this quantum state is showed as\cite{Y1,EKS},
\begin{eqnarray}
C_1=\frac {\la \Psi |\eta ^{\dagger }_k\eta _{l}|\Psi \ra }
{\la \Psi |\Psi \ra }=
\frac {N(L-N)}{L(L-1)},~~k\not= l
\label{correlate}
\end{eqnarray} 
We can find the off-diagonal element is constant for
large distances $|k-l|$. In the thermodynamic limit,
$N,L\rightarrow \infty $ where $N/L=n$, the 
off-diagonal correlation is $n(1-n)$ which generally does not
vanish except for $n=0,1$. 
For the $\eta $-pairing half-filled case
$n=1/2$, this correlation
achieve the maximum.

It is of interest to study the entanglement of the $\eta $-pairing state.
Zanardi $et~al$ \cite{ZW} were the first to
study the pair-wise entanglement of this 
quantum state (\ref{state}).
In the thermodynamic limit, it was showed that
the pair-wise entanglement vanishes. 
This coincides with the entanglement sharing case \cite{KBI}. 
Besides pair-wise entanglement,
other entanglement properties of this
quantum state 
in thermodynamic limit 
should also be studied as those already be done 
for various spin chains.
We will consider the entanglement between $M$ sites and
the rest of the sites. For this case, the entanglement
can be perfectly quantified by the von Neumann entropy of
the reduced density operators of $M$ sites since 
the quantum state (\ref{state}) is a pure state. 

First, let us consider the entanglement between one site with
the rest $L-1$ sites of the quantum state (\ref{state}),
$S_1=-{\rm Tr}\rho _1\log _2\rho _1$,
$S_1$ is the von Neumann entropy of the one site reduced
density operator of the quantum state (\ref{state}). One can find the 
one site reduced density operator takes the form
$\rho _1=(1-\frac NL)|0\ra \la 0|
+\frac NL|1\ra \la 1|$,
where $|0\ra $ is the hole state, $|1\ra $ is the
$\eta $-pair filled state. As previously, we
denote $n=N/L$. The one site entanglement is
\begin{eqnarray}
S_1=-(1-n)\log _2(1-n)-n\log _2n .
\label{onesite}
\end{eqnarray}
So, in the thermodynamic limit, the entanglement between
one site and other sites does not vanish.
Interestingly, for the $\eta $ pair half filled case
$n=1/2$, the entanglement $S_1$ achieve the maximum.
This is the same as the correlation function of
ODLRO. We may also identify that the correlation of
ODLRO in Eq.(\ref{correlate}) has the
hole-$\eta $-pair symmetry, i.e., change $N$ to
$L-N$, this correlation is invariant.
This symmetry also appears in the 
one site entanglement as shown in Eq.(\ref{onesite}). 
In the thermodynamic limit, we can find that
the correlation function of the ODLRO and
the one site entanglement have the same monotonicity 
with respect to the density of the $\eta $-pair $n$,
this can be showed in the Figure 1.

\begin{figure}[h]
\includegraphics[width=6cm]{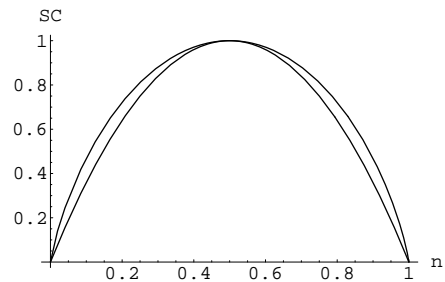}
\caption{The outside curve is $S_1$, the inside curve is $C_1$
normalized by factor 4. For the $\eta$-pair half filled case
$n=1/2$, $S_1$ and $C_1$ achieve the maximal point}
\end{figure}

This result, however, attracts our attention to
the question whether it is possible to quantify 
the entanglement by the correlation function of
ODLRO. This means that correlation function of
ODLRO can be identified as
the entanglement measure of the quantum state
(\ref{state}). 
We know that the widely accepted 
measure of entanglement of pure state is the
von Neumann entropy of the reduced density operator.
However, indeed, we may quantify the entanglement by
other measures, for example, the concurrence
defined by Wootters is also a good and widely accepted
measure of
entanglement\cite{W}. Surprisingly, 
in the thermodynamic limit, 
the concurrence
of one site entanglement corresponding to $S_1$ in (\ref{onesite})
is $n(1-n)$ which is exactly the correlation of the
ODLRO (\ref{correlate}).
Thus the correlation of ODLRO $C_1$ is actually the
concurrence of the quantum state (\ref{state}) between
one site and $L-1$ sites. So, we may say the appearance
of one site entanglement $S_1$ is the hidden reason that
the off-diagonal elements have long-range correlation in
the state (\ref{state}). If the correlation function of
ODLRO is zero, (\ref{state}) is a separable state, 
if it is not zero, (\ref{state}) is an entangled state.

We know that the pair-wise
entanglement of quantum state (\ref{state}) vanishes
in the thermodynamic limit\cite{ZW}. So, the ODLRO does not 
necessarily correspond to the pair-wise entanglement.
However, we find that the correlation function of ODLRO
is the concurrence of 
the one site entanglement of the quantum state (\ref{state}).
The correlation function of
ODLRO showed in Eq.(\ref{correlate}) is in terms of
the pair-wise form, i.e., the correlation function
is concerned about two different sites. 
While the one site entanglement
is in the form: one site with the other
$L-1$ sites. Here we argue that
though the ODLRO is in terms of the pair-wise form, 
since this correlation is the same for all pairs,
it can also be roughly understood as the correlation of one site
with the other $L-1$ sites.

For multipartite state, we may not only consider the
one-site entanglement. The $M$-site entanglement 
is also the basic property of the entanglement.
Next, we consider the entanglement of $M$ sites with
the rest $L-M$ sites of the state (\ref{state}) denoted as
$S_M=-{\rm Tr}\rho _M\log _2\rho _M$,
where $\rho _M$ is the reduced density operator of $M$ sites
of the quantum state (\ref{state}). For convenience, we 
consider the thermodynamic limit, and assume $M$ is finite.
By some calculations, we can find the reduced density
operator of $M$ sites can be represented as
$\rho _M=
%\sum _{i=0}^M
%|\bar {i}\ra \la \bar {i}|\frac {{L-M \choose N-i}
%{M \choose i}}{{L \choose N}}
%\nonumber \\
\sum _{i=0}^M
|\bar {i}\ra \la \bar {i}|f_M(n,i)$,
where we define $f_M(n,i)=n^i(1-n)^{M-i}\frac {M!}{i!(M-i)!}$,
the quantum state $|\bar {i}\ra $ is a symmetric state with
$i$ $\eta$-pairs filled in $M$ sites.
So, we know the von Neumann entropy of $\rho _M$
takes the form
\begin{eqnarray}
S_M=-\sum _{i=0}^Mf_M(n,i)\log _2f_M(n,i).
\label{sm}
\end{eqnarray} 
Here we remark that the $M$ sites entanglement still
has the hole-$\eta$-pair symmetry since mapping   
$|\bar {i}\ra $ to $|\bar {L-i}\ra $ does not change
the von Neumann entropy of $\rho _M$.
As an example, we recover the Eq.(\ref{onesite}) for $M=1$.
%for $M=2$, we have 
%\begin{eqnarray}
%S_2&=&-(1-n)^2\log _2(1-n)^2
%\nonumber \\
%&&-2n(1-n)\log _22n(1-n)
%-n^2\log _2n^2.
%\nonumber 
%\end{eqnarray}
The behavior of $S_M$ are almost the same for all $M$.
$S_M$ achieves the maximum when $n=1/2$ corresponding to
$\eta $-pair half filled case. See Figure 2,3 for the
detail.
\begin{figure}[h]
\includegraphics[width=6cm]{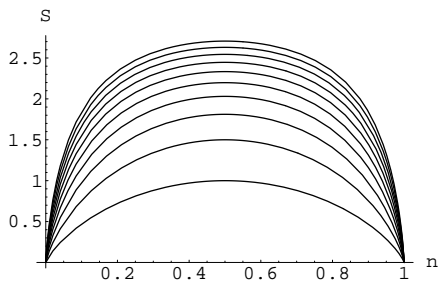}
\caption{The functions of entanglement entropy $S_M$, and we take
$M=1,2,...,10$.}
\end{figure}
\begin{figure}[h]
\includegraphics[width=6cm]{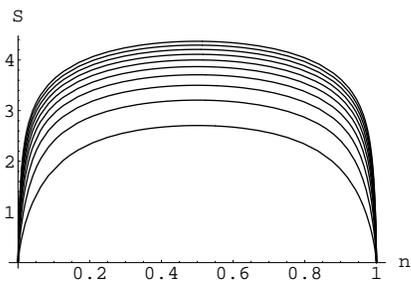}
\caption{The functions of entanglement entropy $S_M$, and we take
$M=10,20,...,100$.}
\end{figure}
We may notice that the quantum state (\ref{state}) is
just symmetric spin state. So, the entanglement of this
quantum state is similar to the symmetric bosonic state in
a lattice. The entanglement of a set of spatial bosonic
modes localized on a graph has been studied in Ref.\cite{Z0}. 
In this Letter, we mainly concern about the relationship
between the entanglement of quantum state (\ref{state})
with the correlation function of the ODLRO.

Our results showed that
the ODLRO in condensed matter physics may
be related with the entanglement. However, we not only
intend to just give an interpretation of ODLRO by
the concurrence, and also we would like to know whether
the quantum information theory can tell us more about 
the phenomenon of ODLRO. 
To study the entanglement property of 
multipartite system, it is natural to consider
not only the one-site entanglement $S_1$, but
also $M$-site entanglement $S_M$. Conversely,
we may wonder whether there exists other types 
of ODLRO in $\eta$-pairing state. For example, we are interested
to know whether the off-diagonal elements of
${\la \Psi |(\eta ^{\dagger }_{k_1}\eta ^{\dagger }_{k_2})
(\eta _{l_1}\eta _{l_2})|\Psi \ra }
/{\la \Psi |\Psi \ra }$ still has the
long-range correlation. This is a natural
question if we relate the ODLRO with
the entanglement. Of course, the pair-wise
correlation as presented in Eq.(\ref{correlate}) is
enough to show that the quantum state (\ref{state})
possesses ODLRO. However, other types of ODLRO may
also be an interesting property of the quantum system.     
Since that the entanglement $S_M$ does not vanish
in the thermodynamic limit, we expect that the
general ODLRO also exist for $\eta $-pairing state.
Next, we consider the general off-diagonal elements 
of ${\la \Psi |
(\eta ^{\dagger }_{k_1}...
\eta ^{\dagger }_{k_M})
(\eta _{l_1}... \eta _{l_M})
|\Psi \ra }$ normalized by 
${\la \Psi |\Psi \ra }$.
Here, for convenience, we still assume $M$ is finite and
$L, N$ are large enough to take the thermodynamic limit.
We also assume that all $l_i$ and $k_j$ are different.

It can be checked that we have the relation
$\la \Psi |\Psi \ra =N!L(L-1)...(L-N+1)$.
By some calculations, we can also find that
\begin{eqnarray}
&&\la \Psi|(\eta ^{\dagger }_{k_1}...\eta ^{\dagger }_{k_M})
(\eta _{l_1}...\eta _{l_M}) |\Psi \ra
\nonumber \\
&=&N^2\la \tilde {\Psi }|(\eta ^{\dagger }_{k_2}...\eta ^{\dagger }_{k_M})
(\eta _{l_2}...\eta _{l_M}) |\tilde {\Psi }\ra ,
\nonumber 
\end{eqnarray}
where state $|\tilde {\Psi }\ra =
(\sum _{j\not=l_1,k_1}\eta _j^{\dagger })^{N-1}|vac\ra $.
With these results, we can readily show that
\begin{eqnarray}
&&C_M=\frac {\la \Psi |
(\eta ^{\dagger }_{k_1}...
\eta ^{\dagger }_{k_M})
(\eta _{l_1}... \eta _{l_M})
|\Psi \ra }{\la \Psi |\Psi \ra }
\nonumber \\
&=&\frac {N...(N-M+1)(L-N)...(L-N-M+1)}
{L...(L-2M+1)}.
\nonumber
\end{eqnarray}
The correlation function $C_M$ does not 
depend on the distances of $|k_i-l_j|, 
i,j=1,...,M$.
Really, we can find other types of ODLRO for the
state (\ref{state}). We can still observe the
hole-$\eta $-pair symmetry since the correlation
function is invariant if we change $N$ to $L-N$.
In the thermodynamic limit, this correlation 
becomes $[n(1-n)]^M$ which is the $M$ power of
the original pair-wise ODLRO.
In this sense, these ODLROs are also related
with the concurrence of the one-site entanglement.

The concurrence of entanglement is well defined 
for 2-level quantum systems\cite{W}. However, there
are no consensus definition of concurrence for higher-level 
quantum systems even for pure states. 
Nevertheless, we remark that the concurrence hierarchy
which include several quantities
are a much general definition for the concurrence\cite{F1}. 
Considering the $M$-site
reduced density operator $\rho _M$, we would like to
point out that the quantity $\prod _{i=0}^Mf_M(n,i)$ which
is equal to $[n(1-n)]^{M(M+1)/2}$ up to a constant factor
is one generalized concurrence. Remind that the general
correlation function of ODLRO is $[n(1-n)]^M$, this
provides us another evidence that ODLRO is closely related
with the entanglement for the $\eta$-pairing state (\ref{state}).

The entanglement of $S_M$ concerns about the
correlation of $M$ sites with the rest $L-M$ sites. 
Comparatively, it
is meaningless to consider the correlation of
the form 
${\la \Psi |(\prod _{i=1}^M\eta ^{\dagger }_{k_i})
(\prod _{j=1}^{M'}\eta _{l_j})
|\Psi \ra }, M\not= M'$ which is actually zero.
So, the definitions of ODLRO and quantum entanglement
cannot be completely identified. However, as we already showed,
they are closely related. 
It is of interest to study other
quantum systems which possess ODLRO. 
The importance of the concept of ODLRO in condensed matter
physics is, as Yang argued that it cannot be
described in classical mechanical terms\cite{Y}.
In the same time,  quantum entanglement
also has no classical analog. In this sense,
if ODLRO cannot be identified
with entanglement in some other systems different from
the $\eta$-pairing state, it is possible that ODLRO and
quantum entanglement describe different aspects of the quantum
systems.  

It is showed that the entropy of entanglement of the ground states of 
several spin chains and one-dimensional Hubbard model 
demonstrate universal scaling behaviour which is related
with the universal properties of the quantum phase transition
\cite{OAFF,ON,VLRK}. We next show numerically that
$S_M$ also obeys universal scaling laws. 
Since $\eta $-pairing state which is an eigenstate of the
three-dimensional Hubbard model is simple, it is 
straightforward to check numerically the scaling of $S_M$
up to, say, $M=3000$ sites.
We obtain the scaling form of $S_M$ as
\begin{eqnarray}
S_M\approx \frac {1}{2}\log _2(M)+k(n),
\label{scale}
\end{eqnarray}
where $k(n)$ depends only on the $\eta $-pair density $n$.
The correspondence
between $k(n)$ and $n$ are presented in
the Figure 4, the exact data are also available. 
The factor $1/2$ in (\ref{scale}) could be related with the central charge
of the conformal field theory\cite{BPZ} as for XY model 
and other models \cite{VLRK,K}.
For cases $n=0$ and $n=1$, we have $S_M=0$, thus the
scaling relation (\ref{scale}) 
does not hold. We checked numerically that near
$n=0$, say, around $n=0.001$, the scaling relation (\ref{scale})
is still correct with high precision. Even near $n=10^{-4}$,
the scaling relation is roughly correct. We remark that
with $n\not =0,1$, the quantum state (\ref{state}) 
is entangled with
ODLRO and is thus in superconducting phase. There will be a
phase transition near $n=0,1$. Our results show that
the scaling relation (\ref{scale}) for
$\eta$-pairing state is general correct except for $n=0,1$. 
Finally, we remark that our result of $S_M$ in (\ref{sm}) is rigorous and
exact.

\begin{figure}[h]
\includegraphics[width=6cm]{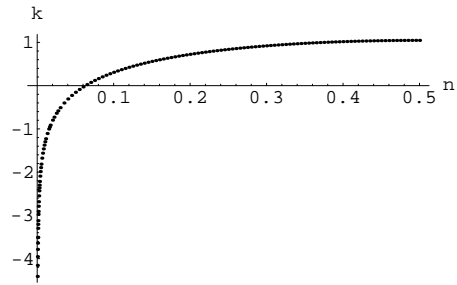}
\caption{The dependence of $k(n)$ on $n$, the values of $k(n)$
are optimized for $M=800$ to satisfy the
scaling relation (\ref{scale}). Since $S_M$ has $\eta$-pair-hole
symmetry, i.e., $S_M$ is invariant for map $n\rightarrow 1-n$, 
we just need to let $n$ ranging from 0 to 0.5.}
\end{figure}

We summarize that quantum entanglement 
plays an important role in the ODLRO of
the quantum state (\ref{state}) which is not tied
to any specific model. Thus entanglement
could also be regarded as one characteristic of
the superconductivity at least for the $\eta $-pairing
quantum state. Though further research
about the relationship between entanglement and
ODLRO is necessary, we already showed that for the well-known
$\eta $-pairing state, the ODLRO and the entanglement ($C_1$ and $S_1$),
two important concepts in condensed matter physics and quantum
information processing, can be identified. 
And further, from the quantum information theory, we 
expected from the fact that $S_M$ in general does 
not vanish that other types of ODLROs exist in thermodynamic limit.
As we already see that the non-zero quantities $C_M$ do exist
in thermodynamic limit. 

{\it Acknowledgements}: The authors would like to thank P.Zanardi 
for useful comments.

\end{document}